\documentstyle[12pt,epsf]{article}
\renewcommand{\thefootnote}{\fnsymbol{footnote}}
\newcommand{\be}{\begin{equation}}
\newcommand{\ee}{\end{equation}}
\newcommand{\ba}{\begin{eqnarray}}
\newcommand{\ea}{\end{eqnarray}}

\def\lsim{\raise0.3ex\hbox{$<$\kern-0.75em\raise-1.1ex\hbox{$\sim$}}}
\def\gsim{\raise0.3ex\hbox{$>$\kern-0.75em\raise-1.1ex\hbox{$\sim$}}}

\begin{document}

\sloppy
\begin{titlepage}

\begin{flushright}
PURD-TH-96-05 \\
CERN-TH/96-167\\
hep-ph/9607386 \\
July  1996
\end{flushright}
\begin{centering}
\vfill
{\large\bf Melting of the Higgs Vacuum: \\
Conserved Numbers at High Temperature}

\vspace{0.4cm}
S. Yu. Khlebnikov$^{~\rm a,}$\footnote{skhleb@physics.purdue.edu}
 and
M. E. Shaposhnikov$^{~\rm b,}$\footnote{mshaposh@nxth04.cern.ch}

\vspace{0.4cm}

{\em $^{\rm a}$ Department of Physics, Purdue University,
West Lafayette, IN 47907, USA}

\vspace{0.1cm}

{\em $^{\rm b}$ Theory Division, CERN, CH-1211 Geneva 23,
Switzerland}

\vspace{5mm}
{\bf Abstract}

\end{centering}

\vspace{0.3cm}\noindent
We discuss the computation of the grand canonical partition sum
describing hot matter in systems with the Higgs mechanism in the
presence of non-zero conserved global charges. We formulate a set of
simple rules for that computation in the high-temperature
approximation in the limit of small chemical potentials. As an
illustration of the use of these rules, we calculate the leading term
in the free energy of the standard model as a function of baryon
number B. We show that this quantity depends continuously on the
Higgs expectation value $\phi$, with a crossover at $\phi\sim T$
where Debye screening overtakes the Higgs mechanism---the Higgs
vacuum ``melts". A number of confusions that exist in the literature
regarding the B dependence of the free energy is clarified.

\vspace{3mm}\noindent

\vfill \vfill
\noindent
PURD-TH-96-05 \\
CERN-TH/96-167\\
July 1996
\end{titlepage}

\noindent
\setcounter{footnote}{0}
\renewcommand{\thefootnote}{\arabic{footnote}}

Screening of gauge fields by the Higgs mechanism is one of the
central ideas both in modern particle physics and in condensed matter
physics. It is often referred to as ``spontaneous breaking" of a
gauge symmetry, although, in the accurate sense of the word, gauge
symmetries cannot be broken in the same way as global ones can. One
might rather say that gauge charges are ``hidden" by the Higgs
mechanism. For example, in the standard model at zero temperature,
particles are characterized by values of the electric charge, but not
the isospin and the weak hypercharge separately.\footnote{ Although
we will use the standard model as an illustration, our results are
not restricted to that case and can apply, for instance, to grand
unified theories.}

At high temperatures, another mechanism for screening becomes
operative---the Debye screening of charges in plasma. The Debye
screening is not associated with ``breaking" or hiding of gauge
charges; on the contrary, it is due to the motion of conserved
charges. Thus, in the standard model, while all elementary
excitations (particles) at zero temperature $(T = 0)$ have only one
conserved gauge charge with respect to electroweak interactions, it
could be different for excitations at $T \not = 0$.

In general, we can envisage a competition between the two screening
mechanisms. The mass of vector bosons generated by the Higgs
mechanism is of order $g\phi$, where $g$ is the gauge coupling and
$\phi$ is the temperature dependent Higgs expectation value. The
electric mass due to the Debye screening is of order $gT$. So, when
$T \ll \phi$, we expect the vacuum classification of particles to be
more or less intact; when $T \gg \phi$, the system forgets that the
gauge symmetry is ``broken".

At $T \sim \phi$ a crossover occurs, which can be called a {\em
melting} of the Higgs vacuum. This regime may or may not be
associated with a phase transition. A gauge theory may have no
gauge-invariant order parameter, and in that case a phase transition
does not have to occur \cite{Fradk79,Banks79}. In fact, it was
recently shown that the electroweak phase transition changes into a
smooth crossover when the mass of the Higgs boson becomes
sufficiently large \cite{notrans}. In this sense, the minimal
standard model is similar to the liquid-gas system: there is no true
distinction between the phases, and the line of first-order phase
transitions ends in a critical point.

The melting of vacuum, associated with a crossover from one screening
mechanism to the other, admits a fully gauge-invariant
characterization using the renormalized average of $\varphi^{\dagger}
\varphi$, where $\varphi$ is the corresponding Higgs field. In the
extreme high-temperature limit, in a weakly coupled theory,
\be
\langle \varphi^{\dagger} \varphi \rangle_R =
\nu\frac{T^2}{12} + {\rm (corrections)},
\ee
where $\nu$ is the number of components of $\varphi$, and the corrections 
are due to interactions. As the temperature lowers, the leading term
starts to deviate from $\nu T^{2}/12$
(at zero temperature it is $v^{2} / 2$).
Temperatures at which the deviation becomes significant (of the order
of the leading term itself) mark the melting crossover.

The above considerations show that the couplings of elementary
excitations to gauge fields (the charges) at high temperature can be
different from those at $T = 0$. This effect can enter calculations
of thermodynamic quantities in systems with the Higgs mechanism,
which contain, in addition, densities of some conserved {\em global}
charges. Therefore, a treatment of such systems requires some care.
As we will see below, the use of kinetic equilibrium requirements in
terms of zero-temperature particle excitations, sometimes invoked
\cite{Harvey,Dolgov:rev,Dreiner} in these circumstances, may, and in
fact does, lead to wrong results at $\phi~ \lsim~ T$.

Although the only application we discuss here will be for the
electroweak sector, it is of some interest to sketch a general
formalism that would allow one to deal with any case when both the
Higgs and the Debye screening mechanisms are present.

In general, a system will have a number of conserved {\em global}
charges, which we denote by $N_i$. These are gauge-invariant
quantities that allow for the introduction of chemical potentials,
$\mu_i$, in the usual way. We thus can define a gauge-invariant grand
partition sum as a Euclidean functional integral:
\be
\exp(-\Omega/T) =
\int {\cal D} \Phi \exp\left[-\int d\tau d^3x (L_E-\sum \mu_i N_i)
\right],
\label{part}
\ee
where $\Phi$ denotes generically all of the fields of our system; the
integral over $\tau$ is restricted to the finite interval from 0 to
$\beta=1/T$, with the usual periodic (antiperiodic) conditions for
bosons (fermions). The Lagrangian $L_E$ includes gauge-fixing terms
and ghosts. The thermodynamic potential $\Omega$ is a gauge-invariant
function of $\mu_i$ and $T$.

Special attention should be paid to integrals over the Euclidean
temporal components of the gauge fields. The integration over their
zero-momentum modes, the set of which will be denoted by ${\cal
A}_4$, enforces the conditions of neutrality of our system with
respect to all gauge charges. These conditions are simply a
consequence of the Gauss constraint, integrated over the whole
three-dimensional (3d) space.\footnote{For definiteness we consider
the 3d space as a compact manifold with large volume which we
eventually send to infinity.}

It is convenient to imagine that the functional integration in
(\ref{part}) is performed in two steps. In the first step, we
integrate out all modes except for ${\cal A}_4$ and the Higgs fields
$\varphi$. Those are integrated out at the second step. The reason
for this two-step procedure is that ${\cal A}_4$ and $\varphi$ can
develop expectation values, which we have to take into account.
Notice that this general method of calculation of ${\tilde \Omega}$
makes no use of the notion of a particle or an elementary excitation
in plasma.

The result of the first step is an effective Euclidean action
${\tilde \Omega}$
\be
\exp(-{\tilde \Omega}/T)=
{\int}' {\cal D} \Phi \exp\left[-\int d\tau d^3x (L_E-\sum \mu_i N_i)
\right],
\label{part'}
\ee
where the prime on the integral shows that the ${\cal A}_4$ and
$\varphi$ integrations are omitted. ${\tilde \Omega}$ is a function
of $\mu_i$, $T$, ${\cal A}_4$ and $\varphi$. One may notice that
$i{\cal A}_4$ play the role of chemical potentials for the
corresponding gauge charges. In general, ${\tilde \Omega}$ is a gauge
non-invariant quantity; the integration over ${\cal A}_4$ and
$\varphi$ will convert it into the gauge-invariant potential
$\Omega$.

Let us now restrict ourselves to weakly-interacting cases when we can
compute ${\tilde \Omega}$ by the loop expansion. We will also assume
that the expectations values of ${\cal A}_4$ and $\varphi$ can be
chosen $x$- and $\tau$- independent. The calculation can be done in
any desired order; here we will need only tree and one-loop terms. At
the tree level, the effective action is the sum of the covariant
derivative term for $\varphi$ and its tree potential,
\be
{\tilde \Omega}_0/{\cal V} = |D_4\varphi|^2 + V_0(\varphi),
\label{tree}
\ee
where in the covariant derivative we set $\partial_4\varphi=0$;
${\cal V}$ is the total 3d volume. One-loop terms can be classified
according to powers of ${\cal A}_4$. Terms independent of ${\cal
A}_4$ add to $V_0$ to produce the usual temperature-dependent
potential for the scalar field. Then, there are linear terms
(tadpoles, see Fig. 1), quadratic terms (Debye masses) as well as
higher powers. We will assume that the chemical potentials for the
global charges are small, $\mu_i/T \ll 1$. In that case, the
expectation values of ${\cal A}_4$ are proportional to $\mu_i$, and
the terms beyond the quadratic ones are suppressed. In addition, the
tadpoles and the Debye masses can be expanded in $\mu_i/T$. For the
tadpoles, the expansion starts with terms linear in $\mu_i$, and for
the Debye masses we can, in the leading order, neglect $\mu_i$
altogether.
\begin{figure}[t]
\vspace*{-4cm}
\hspace{0cm}
\epsfysize=16cm
\centerline{\epsffile{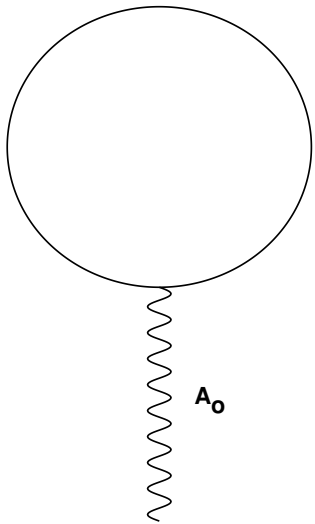}}
\vspace{-6.0cm}
\caption[a]{\protect Tadpole diagram producing terms linear in
chemical potential.}
\end{figure}

A particularly simple set of rules arises in the important case when,
in addition to the above limits, the masses of all particles are much
smaller than $T$. Here the results can be simply formulated
using the spectrum of the high-temperature, ``unbroken" phase.
In that case, the one-loop effective action is obtained as follows.
To each particle species of the high-temperature limit we associate a
chemical potential
\be
\mu_I = \sum_i \mu_i n_i^I + i \sum_j g_j^I {\cal A}_4^j,
\label{chem}
\ee
where $n_i^I$ are the global charges of that species and $g_j$ its
gauge charges; $I$ labels the species. Note that only ${\cal A}_4$,
corresponding to the mutually commuting generators of the gauge
group, should be taken into account. These generators can be
simultaneously diagonalized, and their eigenvalues define what we
mean by the charges $g_j^I$ in eq. (\ref{chem}).\footnote{For the
hypercharge, our convention is $g^I_Y=y_I/2$.} The leading terms in
the one-loop effective action are
\be
{\tilde \Omega}_1/{\cal V} = \left[ V_T(\varphi) - V_0(\varphi) \right]
-T^2 \sum_I \eta_I \mu_I^2
\label{rule}
\ee
where $\eta_I=1/6$ ($\eta_I=1/12$) per single helicity bosonic
(fermionic) particle-antiparticle pair. The complete effective action
needed for our purposes is the sum of the tree action (\ref{tree})
and the one-loop action (\ref{rule}).

As an illustration of the use of these rules, we present here a
calculation of the leading term in the free energy of the standard
model (SM) as a function of the baryon number. It generalizes a
similar calculation done in ref. \cite{ferrer}, where the only
fermionic degrees of freedom were the leptons of the first
generation. The B-dependent part of free energy plays an important
role in the theory of anomalous electroweak B non-conservation. In
our previous calculation of this quantity \cite{ks}, we pointed out
that at sufficiently high temperatures, one may have to use the
spectrum of the ``unbroken" phase even below the phase transition.
Here we extend our previous result and present the leading term in
$F(B)$ for all temperatures $T \gg m_{W}(T)$. In particular, we
demonstrate a crossover in $F(B)$ at $T \sim \phi$.

As is well known, there are $n_f$ strictly conserved global charges
in the SM with massless neutrinos, namely
\be
N_i= \frac{1}{n_f} B -L_i,
\label{conserved}
\ee
where $i=1, ..., n_f$ is the index of fermionic generation. In
addition to them, there are two gauge charges associated with the
underlying SU(2)$\times$U(1) symmetry, which commute with the
Hamiltonian, the global charges defined by eq. (\ref{conserved}), and
each other. These are the hypercharge $Y$,
\be
Y=i\varphi^{\dagger}\stackrel{\leftrightarrow}{D}_0\varphi
+\sum_{i=1}^{n_f}\left[ \frac{1}{3}\bar{Q_i}\gamma_{0}Q_i
+ \frac{4}{3}\bar{U_i}\gamma_{0}U_i
- \frac{2}{3}\bar{D_i}\gamma_{0}D_i
- \bar{L_i}\gamma_{0}L_i
- 2 \bar{E_i}\gamma_{0}E_i\right]
\ee
and the third component of the weak isospin $T_3$,
\be
T_a = \epsilon_{abc}A^b_iG^c_{i0} + \frac{1}{2}
i\varphi^{\dagger}
(\tau^a D_0 - \stackrel{\leftarrow}{D}_0\tau^a)
\varphi
+\frac{1}{2}\sum_{i=1}^{n_f}\left[\bar{Q_i}\gamma_{0}\tau^aQ_i+
\bar{L_i}\gamma_{0}\tau^aL_i\right].
\ee
Here $Q_i$ and $L_i$ are the left quark and lepton doublets,
respectively; $U_i,~D_i$ are the right quark fields, and $E_i$ are
the right leptons.

According to our general rules, we have to introduce $n_f$ chemical
potentials $\mu_i$ to conserved numbers (\ref{conserved}), and
compute the effective action ${\tilde \Omega}(A_4^3,B_4,\mu_i,\phi)$,
where $\varphi^T = (0, \frac{\phi}{\sqrt{2}})$. To make the equations
somewhat more transparent we assume here that the asymmetries in
charges $N_i$ are degenerate in flavour. This assumption is harmless
for the leading-order calculation. It would be inadequate if we
wanted to include corrections in leptonic masses, which are important
in some cosmological scenarios \cite{ks}. This assumption sets all
$\mu_i$ equal to each other and thus replaces them with a single
chemical potential for $B-L$. For future purposes, though, we will
keep separate chemical potentials for $B$ and $L$. The condition of
equilibrium with respect to the anomalous B-non-conservation,
$\mu_B+\mu_L=0$, can be imposed later.

The leading one-loop terms of the high-temperature expansion of the
effective action, for small chemical potentials, are\footnote{We omit
$\phi^3$ term in the effective potential which is not essential for
the present discussion.}
\[
{\tilde \Omega}(A_4^3,B_4,\mu_i,\phi)=
\frac{1}{2}m^2(T) \phi^2 +\frac{1}{4}\lambda \phi^4 +
\]
\be
\frac{1}{8} \phi^2 (g A_4^3-g'B_4)^2+
\frac{1}{2}m_D^2A_4^3A_4^3+\frac{1}{2}m_D'^2B_4B_4
\label{omega}
\ee
\[
-\frac{1}{4}n_f \mu_L^2 T^2 - \frac{1}{9}n_f \mu_B^2 T^2
+i g'B_4 n_f(\frac{1}{3}\mu_L - \frac{1}{9}\mu_B) T^2,
\]
where
\ba
m_D'^2 & = & \biggl(\frac{n_s}{6}+\frac{5n_f}{9}\biggr) g'^2 T^2, \\
m_D^2 & = & \biggl(\frac{2}{3}+\frac{n_s}{6}+\frac{n_f}{3}\biggr) g^2
T^2, \\
m^2(T) & = &  -\frac{1}{2}m_H^2
+T^2\biggl(\frac{1}{2}\lambda+\frac{3}{16}g^2+\frac{1}{16}g'^2+
\frac{1}{4}g_t^2\biggr).
\ea
and $g_t$ is the Yukawa coupling of  $t$-quark.
Terms proportional to $\mu_B^2, \mu_L^2$ are the standard
contributions of the fermionic chemical potentials to $\Omega$, while
the terms linear in chemical potential come from the tadpole diagrams
of Fig. 1.

The equation (\ref{omega}) forms a basis for determining the
equilibrium properties of the hot and dense electroweak plasma at
small chemical potentials. As a first example, let us find the
equilibrium value $B_0$ of the baryonic number at fixed value of the
strictly conserved charge $B-L \equiv X$. That equilibrium value
is obtained from the solution of the system of equations:
\be
\frac{\partial {\tilde \Omega}}{\partial A_4^3}
=\frac{\partial {\tilde \Omega}}{\partial B_4} =0,
\label{zero}
\ee
(neutrality of the system with respect to gauge charges);
\be
-\frac{\partial {\tilde \Omega}}{\partial \mu_B} =B,
{}~~~~~ -\frac{\partial {\tilde \Omega}}{\partial \mu_L} =L
\label{cons}
\ee
(definitions of average baryonic and leptonic charges);
the equilibrium condition
\be
\mu_B+\mu_L=0 ,
\label{equ}
\ee
and the normalization condition
$B-L = X$. We get
\be
B_0=
\frac{4(2 n_f + n_s)m_D^2 + 8(2 + 2 n_f + n_s)m_W^2(\phi)}
{(22 n_f + 13 n_s)m_D^2 + 2(26 + 24 n_f + 13 n_s)m_W^2(\phi)}
(B-L) \; ,
\label{b0}
\ee
where $m_W(\phi)= \frac{1}{2} g \phi$. Even if there is a phase
transition between the symmetric and Higgs phases, in the Higgs phase
with $\phi \ll T$ (the Debye mass is much larger than the
Higgs-induced vector mass), we still get the same result as in the
symmetric phase \cite{ks}
\be
B_0=\frac{4 (2 n_f + n_s)}{22 n_f + 13 n_s} (B-L) \; .
\label{vsmall}
\ee
In the opposite limit $\phi \gg T$ we get
\be
B_0=\frac{4 (2 + 2 n_f + n_s)}{26 + 24 n_f + 13 n_s} (B-L) \; ,
\label{vlarge}
\ee
which coincides with the result of refs.
\cite{Harvey,Dolgov:rev,Dreiner,Ne90}. In the intermediate region,
the value of the baryon number interpolates between the two limiting
cases, see Fig. 2.

\begin{figure}[t]
\vspace*{-5cm}
\hspace{0cm}
\epsfysize=32cm
\centerline{\epsffile{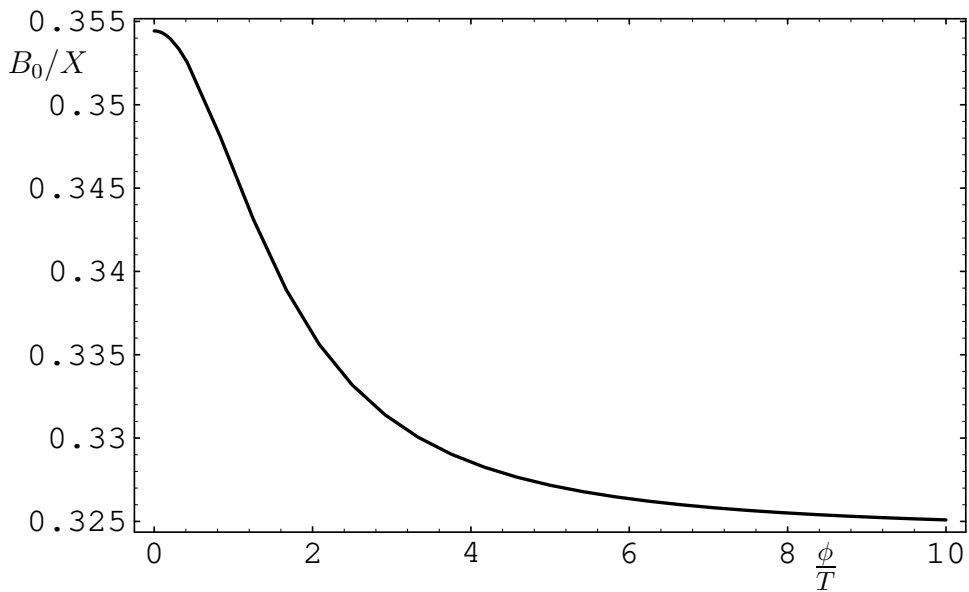}}
\begin{picture}(0,0)
\put(0.9,27.5){$B_0/X$}
\put(11.6,20.8){\large{$\frac{\phi}{T}$}}
\end{picture}
\vspace{-20.5cm}
\caption[a]{\protect Dependence of the equilibrium value of the
baryon number
(in units of $B-L$) on the expectation value of the Higgs field
(in units of temperature) for $n_f=3,~n_s=1$.}
\end{figure}

For cosmological applications, the relation between $B_0$ and $X$
should be taken at the moment of the freeze-out of the sphaleron
processes. For a strongly first-order phase transition, when after
the transition $\phi/T>$1.2--1.5, the freeze-out coincides with the
transition itself, and to a good accuracy one can use the result for
the symmetric phase, eq. (\ref{vsmall}). For a weakly first-order
transition, a second-order transition, or a crossover, one should use
the full eq. (\ref{b0}) with $\phi/T \simeq 1.2$--$1.5$. (That
corresponds to $m_W^2(\phi)/m_D^2\simeq 0.2$--$0.3$.) Looking at Fig.
2, we see that for these values of $\phi/T$, the equilibrium value of
$B$ is approximately in the middle between its two limiting values.

Although the numerical difference between the limiting values is
small, we believe that this calculation is worth-while because it
elucidates the physics responsible for the crossover between the
high- and low- temperature regimes. What matters here is the relation
between $\phi$ and $T$, and not $T$ and $T_c$, as stated in refs.
\cite{Harvey,Dreiner}, or $T$ and $m_W(\phi)$ as stated in ref.
\cite{ks}. The error of the equilibrium reaction analysis of refs.
\cite{Harvey,Dolgov:rev,Dreiner} is in the use of the
zero-temperature spectrum at high temperatures, which led the authors
of these references to conclude that the chemical potential of the
Higgs boson is equal to zero. In our notation, that would correspond
to $g'B_4-g A_4^3=0$. As can be seen directly from eq. (\ref{zero})
with the effective action (\ref{omega}), this condition indeed holds
at $T=0$, but not at any $T\neq 0$.

\begin{figure}[t]
\vspace*{-5cm}
\hspace{0cm}
\epsfysize=32cm
\centerline{\epsffile{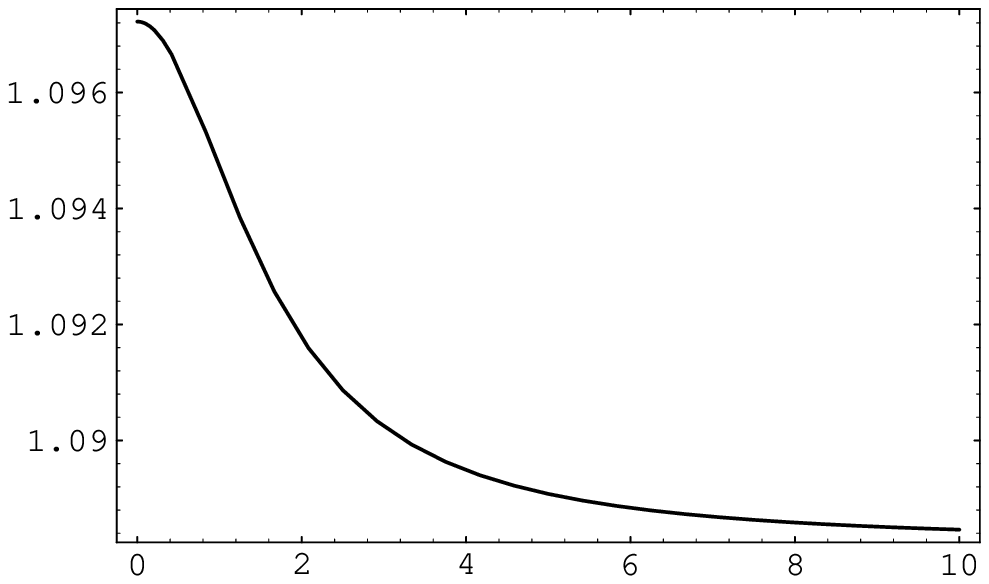}}
\begin{picture}(0,0)
\put(0.3,26.9){\large{$\kappa(\frac{\phi}{T})$}}
\put(11.3,20.0){\large{$\frac{\phi}{T}$}}
\end{picture}
\vspace{-19.5cm}
\caption[a]{\protect Dependence of the coefficient $\kappa$ on the
expectation
value of the Higgs field (in units of temperature) for
$n_f=3,~n_s=1$.}
\end{figure}

Another quantity of interest is the free energy of the system at
fixed baryonic number $B$, not equal to $B_0$, which is used in
the computation of the baryon erasure rate due to sphalerons
\cite{ks}. It is defined as
\be
F(B) = \Omega + \mu_B B + \mu_L L,
\ee
where baryonic and leptonic chemical potentials are to be found from
eqs. (\ref{zero},\ref{cons}) but the equilibrium condition
(\ref{equ}) is now not imposed. The result is
\be
F(B) = \kappa({\phi\over T}) \frac{(B-B_0)^2}{{\cal V} T^2}
+\rm{const},
\label{kap}
\ee
where
\be
\kappa({\phi\over T})=
\frac{3[(22 n_f + 13 n_s)m_D^2 + 2 (26 + 24 n_f + 13
n_s)m_W^2(\phi)]}
{4 n_f[(5n_f  + 3 n_s)m_D^2+ (12 + 11 n_f + 6 n_s) m_W^2(\phi)]}
\ee
For small $\phi/T$ we have
\be
\kappa=\frac{3(22 n_f + 13 n_s)}{4 n_f(5 n_f  + 3 n_s)},
\ee
which for $n_s=1$ coincides with eq. (4.23) of \cite{ks}.
For large $\phi/T$:
\be
\kappa=\frac{3(26 + 24 n_f + 13 n_s)}{2 n_f (12 + 11 n_f + 6 n_s)}
\ee
The dependence of $\kappa$ on $\phi/T$ is shown in Fig. 3.

In conclusion, we have presented a general way of dealing with
conserved global quantum numbers in finite-temperature gauge theories
with the Higgs mechanism. We have formulated a simple set of rules
for computing the grand canonical partition sum in the
high-temperature limit. Using these techniques, we have computed the
leading term of the high-temperature expansion of the free energy of
the standard model, as a function of baryon number. We have shown
that this quantity depends continuously on the Higgs expectation
value $\phi$. The crossover in the free energy is associated with a
``melting" of the Higgs vacuum at $\phi\sim T$, when the Debye
screening overtakes the Higgs mechanism.

S.K. thanks CERN Theory Division, where part of this work was done,
for hospitality. The work of S.K. was supported in part by the U.S.
Department of Energy under grant DE-FG02-91ER40681 (Task B), by the
National Science Foundation under grant PHY 95-01458, and by the
Alfred P. Sloan Foundation.

\end{document}